\documentclass{PoS}

\title{Resonance saturation at next-to-leading order}

\ShortTitle{Resonance saturation at next-to-leading order}

\author{Ignasi Rosell$^{ab}$
 \thanks{I wish to thank the organizers of EFT09 for the useful and pleasant congress. I want to thank P.~Ruiz-Femen\'\i  a and J.J.~Sanz-Cillero for their comments and suggestions. 
This work is supported in part by the Universidad CEU Cardenal Herrera
(Grant No. PRCEU-UCH20/08), by the Spanish Government (Grant No. FPA2007-60323
and Consolider-Ingenio 2010 No. CSD2007-00042, CPAN)
and by the EU No. MRTN-CT-2006-035482 (FLAVIAnet).}\\
 \llap{$^a$}Departamento de Ciencias F\'\i sicas, Matem\'aticas y de la Computaci\'on, Universidad CEU Cardenal Herrera, c/ Sant Bartomeu 55, E-46115 Alfara del Patriarca, Val\`encia, Spain \\
 \llap{$^b$}  IFIC, Universitat de Val\`encia - CSIC, Apartat de Correus 22085, E-46071 Val\`encia, Spain \\

        E-mail: \email{rosell@uch.ceu.es}}

\abstract{A proper estimation of the chiral low-energy constants of Chiral Perturbation Theory is a very
important task. To this end resonance chiral Lagrangians have been used fruitfully. We have studied
the determination of chiral couplings at next-to-leading (NLO) order in the $1/N_C$ expansion, keeping full
control of the renormalization scale dependence. We find that, by imposing short-distance constraints
coming from QCD, resonance saturation at NLO in $1/N_C$ is satisfied. In other words, the chiral couplings can be written 
in terms of the resonance masses and couplings and do not depend explicitly on the coefficients of the chiral operators in 
the Goldstone boson sector of Resonance Chiral Theory.}

\FullConference{International Workshop on Effective Field Theories: from the pion to the upsilon\\

                 2-6 February 2009\\

                 Val\`encia, Spain}

\newcommand{\ket}{\,\rangle}
\newcommand{\bra}{\langle \,}
\newcommand{\cO}{{\cal O}}
\newcommand{\good}{\scriptsize\mbox{`good'}}
\newcommand{\bad}{\scriptsize\mbox{`bad'}}

\begin{document}

\section{Motivation}

Chiral Perturbation Theory ($\chi$PT) is the effective field theory of QCD at very-low energies~\cite{ChPT}. At the moment one needs to include next-to-leading order (NLO) and next-to-next-to-leading order (NNLO) corrections. Therefore, a proper estimation of the chiral low-energy constants (LECs) is a very appealing task. Note the remarkable uncertainties in the phenomenological estimation of the $\mathcal{O}(p^4)$ couplings and the huge number of couplings in the $\mathcal{O}(p^6)$ case, which is a handicap for phenomenological determinations.

Different theoretical approaches to determine the chiral couplings can be used (see for instance  Ref.~\cite{Necco} for the state of the art in the context of lattice QCD or the recent work~\cite{Prades} in the framework of QCD sum-rules).  Here we estimate the LECs by using large-$N_C$ resonance Lagrangians, a fruitful method that has been used commonly. Resonance Chiral Theory (R$\chi$T) is an effective approach to the resonance region~\cite{RChTa,RChTc}. One starts from a phenomenological Lagrangian, including all terms consistent with assumed symmetry principles, which is ruled by the $1/N_C$ expansion. Imposing a good short-distance behavior is one of the main ingredients of our framework, since this matching between functions evaluated with R$\chi$T and QCD allows to reduce the number of unknown parameters. The determination of the chiral couplings by using resonance Lagrangians relies on the assumption that the most important contribution to the chiral couplings come from the physics of the low-lying resonances.
Note that we use Resonance Chiral Theory as a bridge between QCD and ChPT, allowing the determination of LECs in terms of a few parameters.

Resonance large-$N_C$ estimations of chiral couplings are usual, both at $\mathcal{O}(p^4)$~\cite{RChTa} and at $\mathcal{O}(p^6)$ \cite{RChTc}. Indeed, this is the usual way to estimate the new parameters appearing in two-loop calculations. Recently, determinations of low-energy constants of $\chi$PT at the next-to-leading order in the $1/N_C$ expansion have been developed~\cite{cata,juanjo}, which of course reduce the uncertainty. Considering that the dependence of the LECs with the renormalization scale is a subleading effect in $1/N_C$, the usual large-$N_C$ estimations are unable to control the renormalization-scale dependence (typically one assumes that LO estimations correspond to a value $\mu_0=M_\rho$), which can be a sizable effect.

Obviously, the resonance estimation of any $\chi$PT constant $L_i$ depends on the equivalent R$\chi$T constant $\widetilde{L}_i$, corresponding to the coupling related to the same operator, but in the theory where the resonances are still active degrees of freedom. In Ref.~\cite{RChTa} it was found that, at leading-order and for $\mathcal{O}(p^4)$ LECs, $\widetilde{L}_i$ vanish in the antisymmetric formalism and are fixed in the Proca formalism, in both cases due to short-distance constraints. The main aim of this work is to prove that this effect still holds at subleading order in $1/N_C$~\cite{PRD2}, closing the analyses started in Refs.~\cite{oneloop,PRD}.

\section{The framework}

\subsection{Chiral Perturbation Theory}

The chiral symmetry constraints encoded in Chiral Perturbation Theory provide a perturbative expansion in powers of light quark masses and momenta~\cite{ChPT}, so that the effective lagrangian is organized following this expansion,
\begin{eqnarray}
 \mathcal{L}_{\chi PT} &=& {\displaystyle \sum_{n \geqslant 1}} \mathcal{L}_{2n}^{\chi PT} \,. \label{chptlagrangian}
\end{eqnarray}
The leading-order term
\begin{eqnarray}
 \mathcal{L}_2^{\chi PT} &=& \frac{F^2}{4} \bra u_\mu u^\mu + \chi_+ \ket \,
\end{eqnarray}
contains only two couplings, the meson decay constant in the chiral limit $F$ and the constant $B_0$ appearing in $\chi$, which is related to the quark condensate. These parameters cannot be obtained from the underlying theory. Higher orders couplings collect information from degrees of freedom that have been integrated out and therefore they can be estimated from high-energy scales. Moreover, the number of couplings increases fast with the order,
\begin{equation}
 \mathcal{L}_4^{\chi PT}\,=\,{\displaystyle \sum_{i=1}^{10}} L_i \cO_i^{(4)}\,, \qquad \mathcal{L}_6^{\chi PT} \,=\, {\displaystyle \sum_{i=1}^{90}} C_i \cO_i^{(6)} \,,
\end{equation}
where $\cO_i^{(2n)}$ are operators of $\cO(p^{2n})$ in the chiral expansion. In order to relate the $\cO(p^4)$ coupling with different processes, it is quite convenient to show explicitly the expression of the NLO piece,
\begin{eqnarray} \label{ChPTp4}
\mathcal{L}_4^{\chi PT}\!\!&=\!\!& L_1 \bra u_\mu u^\mu \ket^2 \!+\! L_2 \bra u_\mu u^\nu \ket \bra u^\mu u_\nu \ket \!+\! L_3 \bra u_\mu u^\mu u_\nu u^\nu \ket + L_4 \bra u_\mu u^\mu \ket \bra \chi_+ \ket 
  \!+\!  L_5 \bra u_\mu u^\mu \chi_+ \ket  \\ && \!+\! L_6 \bra \chi_+ \ket^2 \nonumber \!+\! L_7 \bra \chi_-\ket^2
\!+\! L_8/2 \, \bra \chi_+^2 + \chi_-^2 \ket  
\!-\!i L_9 \bra f^{\mu\nu}_+ u_\mu u_\nu \ket \nonumber  \!+\!L_{10}/4 \, \bra f_{+ \mu\nu}f_+^{\mu\nu}-f_{-\mu\nu}f_-^{\mu\nu}\ket  \,,
\end{eqnarray}
where the $SU(3)$ case is considered and we have dismissed contact terms and operators that vanish when the equations of motion are used. Since the vector, axial-vector, scalar and pseudoscalar sources are contained in the chiral tensors $f_{+}^{\mu \nu},\,f_{-}^{\mu \nu},\,\chi_{+}$ and $\chi_{-}$, respectively, and $u_\mu$ involves at least one Goldstone boson, it follows that at $\cO(p^4)$ in the chiral limit:
(i) $L_1$, $L_2$ and $L_3$ determine the Goldstone boson scattering,
(ii) $L_4$ and $L_5$ the scalar form factor of the pion,
(iii) $L_6+L_7$ and $L_8$ the difference of the scalar and pseudoscalar correlators,
(iv) $L_6$ the two-point Green function of two scalar densities $\bar{q} q$  and  $\bar{q}' q'$   with $q\neq q'$,
(v) $L_9$ the vector form factor of the pion,
and (vi) $L_{10}$ the difference of the two-point correlation function of  vector and axial-vector currents.

\subsection{Resonance Chiral Theory}

The main problems to develop a formal effective theory in the resonance region, $M_\rho \lesssim E \lesssim 2$~GeV, are the existence of many resonances with close masses and the absence of a natural expansion parameter. Large-$N_C$ QCD furnishes a practical scenario to work with. The limit of an infinite number of quark colors turns out to be a very useful instrument to understand many features of QCD and supplies an alternative power counting to describe the meson interactions~\cite{largeNC}. Tree-level interactions between an infinite spectrum of narrow states implemented in a chiral invariant lagrangian provide the LO ($N_C \rightarrow \infty$) contribution to Green functions of QCD currents, being the NLO corrections given by one-loop diagrams. The model-dependence of this description is the cut of the tower of resonances, which is supposed to be a good assumption since contributions from higher states are assumed to be suppressed by their masses; moreover the approximation is supported by the phenomenology. In fact, the truncation of the tower and the choice of an appropriate set of short-distance constraints for each case constitute the so-called minimal hadronic approximation~\cite{MHA}, which can be implemented in an equivalent way by using meromorphic approximations~\cite{MHA,GF} or a chiral resonance lagrangian~\cite{RchTGreen}. Some issues related to the truncation of the spectrum to a finite number of resonances are discussed in Refs.~\cite{truncation}.

The lagrangian of Resonance Chiral Theory can be organized according to the number of resonance fields,
\begin{eqnarray}
 \mathcal{L}_{R\chi T}&=& \mathcal{L}^{GB} + \mathcal{L}^{R_i} + \mathcal{L}^{R_iR_j} + \mathcal{L}^{R_iR_jR_k} + \dots \,, \label{rchtlagrangian}
\end{eqnarray}
 where $R_i$ stands for resonance multiplets of vectors $V(1^{--})$, axial-vectors $A(1^{++})$, scalars $S(0^{++})$ and pseudoscalars $P(0^{-+})$. Interactions with large number of derivatives in Eq.~(\ref{rchtlagrangian}) tend to violate the QCD ruled asymptotic behavior of Green Functions or form factors. It is important to distinguish between $\mathcal{L}_{\chi PT}$ and $\mathcal{L}^{GB}$: although both have the same structure and operators they correspond to different theories and consequently the values of the couplings are different, {\it i.e.} for instance $L_i \neq \widetilde{L}_i$. The truncation of the infinite tower of resonances of the large-$N_C$ spectrum to the lowest-lying multiplets is not essential in what follows, but can be assumed to ease the discussion. The second term in Eq.~(\ref{rchtlagrangian}) reads,
\begin{eqnarray}
&\qquad \qquad \qquad \qquad \qquad \mathcal{L}^{R_i} = \mathcal{L}^V + \mathcal{L}^A + \mathcal{L}^S + \mathcal{L}^P \, ,  &\phantom{\frac{1}{2}} \nonumber \\
&\mathcal{L}^V_{(2)} = \displaystyle\frac{F_V}{2\sqrt{2}} \bra V_{\mu\nu} f^{\mu\nu}_+ \ket \,+\, \frac{i\, G_V}{2\sqrt{2}} \bra V_{\mu\nu} [u^\mu, u^\nu] \ket \, ,  \qquad
&\mathcal{L}^A_{(2)} = \frac{F_A}{2\sqrt{2}} \bra A_{\mu\nu} f^{\mu\nu}_- \ket\, , \nonumber\\
&\mathcal{L}^S_{(2)} = c_d \bra S u_\mu u^\mu\ket\,+\,c_m\bra S\chi_+\ket\,  \, , \phantom{\frac{1}{2}}   \qquad
&\mathcal{L}^P_{(2)} = i\,d_m \bra P \chi_- \ket \phantom{\frac{1}{2}} \, , \label{1Rlagrangian}
\end{eqnarray}
where only terms with the minimum number of derivatives have been shown. It is convenient to remark that high-energy constraints give relations between these couplings. 

\section{Resonance saturation}

The purpose of this work is to advance in the comprehension of how the low-energy couplings of the lagrangian of Eq.~(\ref{chptlagrangian}) are estimated by integrating out the resonance fields in the resonance lagrangian of Eq.~(\ref{rchtlagrangian}). In particular, we want to understand the role of the couplings of $\mathcal{L}^{GB}$. Upon integration of the resonances one gets an expression of any chiral coupling in terms of the parameters in the R$\chi$T Lagrangian:
\begin{equation}
 L_i(\mu)\,=\,\widetilde{L}_i(\mu) + f_i(M_{R},\alpha_R;\mu) \,, \qquad  \quad C_i(\mu)\,=\,\widetilde{C}_i(\mu) + g_i(M_{R},\alpha_R;\mu)\,,\label{saturation}
\end{equation}
where $f_i(M_{R},\alpha_{R};\mu)$ and $g_i(M_{R},\alpha_{R};\mu)$ are the contribution stemming from the low-energy expansion of the resonance contributions, being $M_R$ the resonance masses and $\alpha_R$ any R$\chi$T coupling accompanying operators with resonances. A convenient definition of resonance saturation is that $\widetilde{L}_i(\mu)$ can be fixed completely after the matching procedure and then $L_i(\mu)$ are given as functions of only $M_R$ and $\alpha_R$. See that with this definition the saturation is accomplished for any value of $\mu$ (the ``extreme'' version of resonance saturation pointed out in Ref.~\cite{cata}).

\subsection{Leading-order}

At leading-order in $1/N_C$ and focusing on the LECs of Eq.~(\ref{ChPTp4}), in Refs.~\cite{RChTa} it was found that $\widetilde{L}_i$ vanish due to short-distance constraints and Eq.~(\ref{saturation}) turns out to be
\begin{eqnarray}
L_1=\frac{G_V^2}{8M_V^2}\,,\quad  
L_2=\frac{G_V^2}{4M_V^2}\,, \quad
L_3=-\frac{3G_V^2}{4M_V^2}\,+\,\frac{c_d^2}{2M_S^2}\, ,\quad
L_4=L_6=L_7=0\,,  \quad
L_5=\frac{c_dc_m}{M_S^2}\,,  \nonumber \\
L_8=\frac{c_m^2}{2M_S^2}\,-\,\frac{d_m^2}{2M_P^2} \,,  \quad
L_9=\frac{F_VG_V}{2M_V^2}\,, \quad 
L_{10}=-\frac{F_V^2}{4M_V^2}\,+\,\frac{F_A^2}{4M_A^2}\,,\qquad
\label{rescont}
\end{eqnarray}
that is, one has been able to determine the $\cO(p^4)$ chiral couplings of Eq.~(\ref{ChPTp4}) in terms of the resonance parameters of Eq.~(\ref{1Rlagrangian}). For higher-order low-energy constants the same is supposed to arise.  

\subsection{Next-to-leading order}

Resonance saturation at NLO with a resonance lagrangian involving only scalars and pseudoscalars mesons was discussed in Ref.~\cite{PRD}. Ref.~\cite{PRD2} addresses the more general case which accounts also for spin-1 resonance fields. 

We analyze in what follows the large-$q^2$ structure of the two-current correlators, pion form factors and Goldstone scattering amplitude separately~\cite{PRD2}:
\begin{enumerate}
\item $\Pi^{\mathrm{1-loop}}(q^2)$: the difference of the two-point functions built from two scalar ($SS$) and pseudoscalar ($PP$) densities ($\Pi_{SS-PP}(q^2)$), or from two vector ($VV$) and axial-vector ($AA$) currents ($\Pi_{VV-AA}(q^2)$). As it has been pointed out these amplitudes are related in the chiral limit to the estimation of the $L_{6-8}$ and $L_{10}$ chiral $\cO(p^4)$ couplings.
\item $\mathcal{F}^{\mathrm{1-loop}}(q^2)$: the scalar and vector form factor of the pion. These amplitudes are related to $L_{4-5}$ and $L_9$.
\item $T^{\rm 1-loop}(\nu,t=0)$: the forward scattering amplitude of $s\leftrightarrow u$ symmetric amplitudes, with $\nu\equiv (s-u)/2$. These amplitudes are related to $L_{1-3}$.
\end{enumerate} 
 To simplify we will consider only resonance operators in ${\cal L}^{R_i},\,{\cal L}^{R_i R_j}\,\dots$ with a chiral tensor up to ${\cal O}(p^2)$. The generalization of our findings for the case of higher-order interaction terms would be straightforward.

After reduction to scalar integrals, all one-loop terms are proportional to scalar two- and one- point functions in the correlators; three-, two- and one- point functions in the form factors; and four-, three-, two- and one- point functions in the scattering amplitudes. Expanding out the expressions for $q^2\rightarrow \infty$ and taking into account that we allow spin-$1$ mesons in the absorptive part, the one-loop amplitudes have the form: 
\begin{eqnarray}
\Pi^{\mathrm{1-loop}}(q^2) &=& \left( \hat{\lambda}^{(0)} q^0 + \hat{\lambda}^{(2)}q^2+\hat{\lambda}^{(4)}q^4    \right) \log{\frac{-q^2}{M_R^2}}
+  \left( \hat{\gamma}^{(0)}  q^0 + \hat{\gamma}^{(2)} q^2  + \hat{\gamma}^{(4)} q^4   \right) +\cO\left(\!\frac{1}{q^2}\!\right)  , \nonumber \\
\mathcal{F}^{\mathrm{1-loop}}(q^2) &=& \left( \hat{\kappa}^{(2)}q^2\right) \log^2{\frac{-q^2}{M_R^2}}
+ \left( \hat{\lambda}^{(2)} q^2 + \hat{\lambda}^{(4)}q^4+\hat{\lambda}^{(6)}q^6 +\hat{\lambda}^{(8)}q^8    \right)  \log{\frac{-q^2}{M_R^2}}\,
\nonumber \\
&& \qquad 
+ \left( \hat{\gamma}^{(2)}  q^2 + \hat{\gamma}^{(4)} q^4  + \hat{\gamma}^{(6)} q^6 + \hat{\gamma}^{(8)} q^8   \right)
+\cO\left(q^0\right)  , \nonumber \\
T^{\rm 1-loop}(\nu,0) &=&  \left(\hat{\kappa}^{(4)}\nu^2+\hat{\kappa}^{(8)} \nu^4 + \hat{\kappa}^{(12)}\nu^6 \right) \log^2{\frac{-\nu^2}{M_R^4}}
+ \left( \hat{\lambda}^{(4)} \nu^2 + \hat{\lambda}^{(8)}\nu^4+\hat{\lambda}^{(12)}\nu^6    \right)  \log{\frac{-\nu^2}{M_R^4}}
\nonumber \\
&& \qquad  + \left(   \hat{\gamma}^{(4)} \nu^2  + \hat{\gamma}^{(8)} \nu^4 + \hat{\gamma}^{(12)} \nu^6   \right) +\mathcal{O}\left(\nu^0\right)  , \label{oneloop}
\end{eqnarray}
with $M_R$ some arbitrary mass scale chosen to make the argument of the logarithms dimensionless and $\hat{\kappa}^{(n)}$, $\hat{\lambda}^{(n)}$ and $\hat{\gamma}^{(n)}$ combinations of resonance parameters. Note that for instance $\hat{\gamma}^{(4)}$ in the correlators differ from $\hat{\gamma}^{(4)}$ in the form factors or scattering amplitudes (we are using the same notation for simplicity).

Local terms from $\mathcal{L}^{\mathrm{GB}}$ also contribute to the amplitudes through a polynomial in the $\widetilde{L}_i,\,\widetilde{C}_i\dots$ couplings:
\begin{equation}
\Pi^{\rm{GB}}(q^2) =  \widetilde{L}_J +   \widetilde{C}_J \,q^2 + \dots \,, \quad
\mathcal{F}^{\mathrm{GB}}(q^2) = \frac{\widetilde{L}_J\, q^2}{F^2} + \frac{\widetilde{C}_J \, q^4}{F^2} + \dots  \,, \quad
T^{\mathrm{GB}}(\nu,0) = \frac{\widetilde{L}_J\,  \nu^2}{F^4} + \dots \,,  \label{GB}
\end{equation}
where the $\widetilde{L}_J$, $\widetilde{C}_J$... refer to corresponding LECs or combination of them for the amplitude.

Now one has to consider the short-distance constraints coming from QCD: the studied combination of correlators are supposed to vanish for $q^2\rightarrow \infty$~\cite{PRD2,VA}; the Brodsky-Lepage rules for the form factors state that the pion form factors behave at worst as a constant for large momentum transfer~\cite{BrodskyLepage}; and the behavior of the forward scattering amplitude $T\left(\nu\rightarrow \infty,0\right)\sim \nu^0$ at high energies. These requirements translate into conditions on the terms shown in Eqs.~(\ref{oneloop}) and (\ref{GB}), which have the wrong high-energy behavior. Because of their different analytical structure, the cancellations must occur separately for the logarithmic and polynomial parts. The vanishing of the non-polynomial  part requires that $\hat{\kappa}^{(n)}=\hat{\lambda}^{(n)}=0$. The cancellation of the remaining polynomial is then achieved by tuning the local contributions from $\mathcal{L}^{\rm GB}$ to fulfill the equations
\begin{eqnarray}
\widetilde{L}_J\,  +   \hat{\gamma}^{(0)}  =0 \, ,
\qquad
 \widetilde{C}_J\,  +   \hat{\gamma}^{(2)}  = 0 \, ,\quad \dots
 \label{LGBfixing}
\end{eqnarray}
in the case of the correlators. The extrapolation to the form factors and scattering amplitudes is straightforward. These constraints  fix the value of the
corresponding ${\cal L}^{\mathrm{GB}}$ couplings, that is the saturation of the LECs holds at NLO in $1/N_C$. 

\section{Summary}

Chiral Perturbation Theory is the effective field theory of QCD at very low energies and its further progress depends on our ability to estimate the low-energy constants. It is well-known that the couplings of every effective field theory collect information from degrees of freedom that have been integrated out to obtain the low-energy lagrangian. Accordingly chiral couplings would receive an important contribution from the low-lying resonances that do not appear in $\chi$PT. Resonance Chiral Theory is an appropriate framework to incorporate the massive mesonic states. This phenomenological approach makes use of the $1/N_C$ expansion and one of its main ingredients is the employment of the short-distance constraints prescribed by QCD. 

Resonance saturation within this formalism can be defined precisely: it states that the $\chi$PT LECs can be written in terms of only the resonance couplings and masses. The statement is not trivially satisfied because the R$\chi$T  amplitudes also depend on the parameters $\widetilde{L}_i$, $\widetilde{C}_i,\dots$ of the Goldstone boson sector which describes the self-interactions of the Goldstone bosons in the presence of resonances.

The determination of the $\chi$PT couplings at the next-to-leading order in $1/N_C$ is an important issue because the dependence of the $\chi$PT couplings with the renormalization scale is a subleading effect in the $1/N_C$
counting. Here we have analyzed the resonance saturation at subleading order and we have found that possible unknown $\cO(p^4)$ (or higher) parameters are determined as soon as one considers short-distance constraints.

\end{document}